# Potential of proteasome inhibitors to inhibit cytokine storm in critical stage COVID-19 patients

Ralf Kircheis[a,*,1], Emanuel Haasbach[b], Daniel Lueftenegger[a,2], Willm T. Heyken[a,3], Matthias Ocker[c,4], and Oliver Planz[b]

[a] Virologik GmbH, Henkestrasse 91, 91052 Erlangen, Germany,

[1] Present address: Syntacoll GmbH, Donaustrasse 24, 93342 Saal a.d. Donau, Germany

[2] Present address: Biogen GmbH, Riedenburger Str. 7, 81677 Munich, Germany

[3] Present address: TÜV SÜD Product Service, Ridlerstraße 65, 80339 Munich, Germany

[b] Institute of cell Biology and Immunology, Eberhard Karls University Tuebingen, Germany

[c] Institute for Surgical Research, Philipps University of Marburg, 35043 Marburg, Germany

[4] Present address: Translational Medicine & Clinical Pharmacology, Boehringer Ingelheim Pharma GmbH, 55216 Ingelheim, Germany and Charité University Medicine Berlin, 10117 Berlin, Germany,

*Corresponding author, current address:   Ralf Kircheis (MD PhD)

*Director Research & Development*

*Syntacoll GmbH*

*Donaustrasse 24*

*93342 Saal a.d. Donau*

*Germany*

*Phone: +49 151 167 90606*

*Mail: rkircheis@syntacoll.de*

**Abstract**

Patients infected with SARS-CoV-2 show a wide spectrum of clinical manifestations ranging from mild febrile illness and cough up to acute respiratory distress syndrome, multiple organ failure and death. Data from patients with severe clinical manifestations compared to patients with mild symptoms indicate that highly dysregulated exuberant inflammatory responses correlate with severity of disease and lethality. Significantly elevated cytokine levels, i.e. cytokine storm, seem to play a central role in severity and lethality in COVID-19. We have previously shown that excessive cytokine release induced by highly pathogenic avian H5N1 influenza A virus was reduced by application of proteasome inhibitors. In the present study we present experimental data of a central cellular pro-inflammatory signal pathways, NF-κB, in the context of published clinical data from COVID-19 patients and develop a hypothesis for a therapeutic approach aiming at the simultaneous inhibition of whole cascades of pro-inflammatory cytokines and chemokines via blocking the nuclear translocation of NF-κB by proteasome inhibitors. The simultaneous inhibition of multiple cytokines/chemokines using clinically approved proteasome inhibitors is expected to have a higher therapeutic potential compared to single target approaches to prevent cascade (i.e. triggering, synergistic, and redundant) effects of multiple induced cytokines and may provide an additional therapeutic option to be explored for treatment of critical stage COVID-19 patients.

1. **Introduction**

Coronaviruses - enveloped single-stranded RNA viruses - are broadly distributed in humans and animals. While most human coronavirus (hCoV) infections show mild symptoms, there are highly pathogenic hCoV, i.e. the severe acute respiratory syndrome virus (SARS-CoV) and the Middle East respiratory syndrome coronavirus (MERS-CoV), with 10% and 37% mortality, respectively. The novel coronavirus SARS-CoV-2 with more than 10 mio infected persons and 500.000 deaths worldwide (https://coronavirus.jhu.edu/) End of June 2020 has become a global pandemic with enormous medical and socio-economic burden. Patients infected with SARS-CoV-2 show a wide spectrum of clinical manifestations ranging from mild febrile illness and cough up to acute respiratory distress syndrome (ARDS), multiple organ failure, and death, *i.e.* a clinical picture in severe cases that is very similar to that seen in SARS-CoV and MERS-CoV infected patients. While younger individuals show predominantly mild-to-moderate clinical symptoms, elderly individuals frequently exhibit severe clinical manifestations (Huang et al., 2020; Xu et al., 2020, Zheng et al., 2020, Wang et al., 2020). Post-mortem analysis showed pronounced Diffuse Alveolar Disease with capillary congestion, cell necrosis, interstitial oedema, platelet-fibrin thrombi, and infiltrates of macrophages and lymphocytes (Carsana et al., 2020). Recently, the induction of endotheliitis in various organs (including lungs, heart, kidney, and intestine) by SARS-CoV-2 infection as a direct consequence of viral involvement and host inflammatory response was shown (Varga et al., 2020).

SARS-CoV-2 binds with its spike (S) protein to the angiotensin-converting enzyme-related carboxypeptidase-2 (ACE-2) receptor on the host cell using the cellular serine protease TMPRSS2 for S protein priming (Hoffmann et al., 2020). The ACE-2 receptor is widely expressed in pulmonary and cardiovascular tissues, hematopoietic cells, including monocytes and macrophages which may explain the broad range of pulmonary and extra-pulmonary effects of SARS-CoV-2 infection including cardiac, gastrointestinal organs, and kidney affection (Varga et al., 2020). The morbidity and mortality of highly pathogenic hCoV is still incompletely understood. Virus-induced cytopathic effects and viral evasion of the host immune response play a role in disease severity. However, clinical data from patients, in particular those with severe clinical manifestations indicate that highly dysregulated exuberant inflammatory and immune responses correlate with severity of disease and lethality (Huang et al., 2020; Carsana et al., 2020; Tay et al., 2020, Schett et al., 2020; Moore et al., 2020). Significantly elevated cytokine levels, *i.e.* cytokine storm, seem to play a central role in severity and lethality in SARS-CoV-2 infections, with elevated plasma levels of IL-1β, IL-7, IL-8, IL-9, IL-10, G-CSF, GM-CSF, IFNγ, IP-10, MCP-1, MIP-1α, MIP-1β, PDGF, TNFα, and VEGF in both, ICU (Intensive care unit) patients and non-ICU patient. Significantly higher plasma levels of IL-2, IL-7, IL-10, G-SCF, IP-10, MCP-1, MIP-1α, and TNFα were found in patients with severe pneumonia developing ARDS and requiring ICU admission and oxygen therapy compared to non-ICU patients showing pneumonia without ADRS (Huang et al., 2020). Interestingly, increased serum levels of pro-inflammatory cytokines and chemokines (e.g. IL-1β, IL-6, IL-12 and IL-8) were associated with pulmonary inflammation and extensive lung damage also in SARS-CoV and MERS-CoV infected patients (Wong et al., 2004; Channappanvar & Perlman, 2017). Moreover, avian H5N1 and H1N1 influenza virus infections with high lethality in humans, showed excessive alveolar inflammatory infiltrates and high levels of pro-inflammatory cytokines and chemokines (Cheung et al., 2008; de Jong et al., 2006; Wong et al., 2006) in human cell lines, mice, and macaques (Chan et al., 2005; Kobasa et al., 2007) and in humans infected with H1N1 (Perez-Padilla et al., 2009).

We previously showed that elevated cytokine release of IL-α/β, IL-6, MIP-1β, and TNF-α induced by highly pathogenic avian H5N1 influenza A virus was reduced by application of the reversibly binding, 20P proteasome inhibitor VL-01 (Leban et al., 2008, Haasbach et al., 2011). The ubiquitin-proteasome system (UPS) is a key player in regulating the intracellular sorting and degradation of proteins, and proteasome inhibitors, in particular Bortezomib (Velcade™), are used for cancer treatment (Adams 2002). Furthermore, VL-01 reduced influenza virus replication in human lung adenocarcinoma cells as demonstrated with various H1N1 and H5N1 virus strains. VL-01-aerosol-treatment of BALB/c mice reduced progeny virus titers in the lung by >90% and enhanced survival of mice after infection with a 5-fold MLD50 of the human influenza A virus strain A/Puerto Rico/8/34 (H1N1) (Haasbach et al., 2011). Notably, the UPS seems to play a role in the coronavirus infection cycle as shown by the effect of proteasome inhibitors on SARS-CoV virus entry (Raaben et al., 2010) and the interaction of the SARS-CoV nucleocapsid protein with the proteasome subunit p42 in host cells (Wang et al., 2010). Proteasome inhibitors were shown to inhibit replication of other RNA viruses, such as HIV (Schubert et al., 2000). The inhibition of the nuclear translocation of the transcription factor NF-κB seems to be central to inhibiting cytokine release and may also be involved in the inhibition of nuclear export of viral RNPs (Haasbach et al., 2011; Wurzer et al., 2003). The general mechanism of NF-κB inhibition by proteasome inhibitors (described by Adams 2002; Moynagh et al., 2005) is mediated by inhibition of the proteasomal degradation of the cytosolic inhibitor IκBα, this way keeping NF-κB sequestered in the cytosol, and preventing the LPS-, cytokine-, or RNA virus-induced (via binding to TLR4, cytokine receptor, or Toll-like receptors, respectively) nuclear translocation of NF-κB and transcription of multiple pro-inflammatory proteins, such as cytokines, chemokines, adhesion molecules and growth factors (Fig. 1). Common to these different signaling pathways is that all of them join into a common downstream signaling sequence triggering proteasomal degradation of IκBα which results in release and translocation of NF-κB into the nucleus (Moynagh et al., 2005) (Fig. 1).

Based on experimental data from *in vitro* and *in vivo* models we present here a hypothesis for a new therapeutic option for treatment of critical stage COVID-19 patients aiming at the simultaneous inhibition of cascades of pro-inflammatory cytokines and chemokines induced during acute viral infection by blocking the nuclear translocation of NF-κB using clinically approved proteasome inhibitors.

## 2. Materials and methods

### 2.1. Proteasome Inhibitors

The novel proteasome inhibitor VL-01 (Z-Trp-Trp-Phe-aminohydantoin, MW = 752.82 g/mol) described by Leban et al. (2008) was synthesized at Almac (Ireland). The proteasome inhibitor Bortezomib was purchased from clinical pharmacy.

### 2.2. Measurement of proteasome activity (P20 assay) in cell lysates

#### 2.2.1. Cell lysate preparation

Jurkat T-cells were lysed in hypotonic P20 Lysis-Buffer (5 mM EDTA, pH 8), freeze-thawed, spun down, and the supernatant, containing proteasomes, mixed with P20 Stabilizing-Buffer (40 mM HEPES, 1 mM EDTA, 20 % Glycerol). The lysates were aliquoted and stored at -80 °C.

#### 2.2.1. P20 Lysate Assay

Lysates were thawed on ice. Serial dilutions of the test compounds in cell lysates were incubated at 37 °C for 1 hour. 50 µl was transferred to a black 96-well plate and 150 µl pre-heated Ys substrate buffer (78 µM Ys substrate, Suc-Leu-Leu-Val-Tyr-AMC (Bachem, Bubendorf, Switzerland), 26 mM HEPES, 0.7 mM EDTA, 0.065 % SDS) added for measurement of the chymotrypsin-like activity. Es (Z-Leu-Leu-Glu-AMC) or Rs (Boc-Leu-Arg-Arg-AMC) substrate (Bachem, Switzerland) were used for measuring the caspase-like and trypsin-like activities, respectively. TP20 activity was measured using a Fluorescence Reader (Synergy HT, Biotek). To normalize the fluorescence data, sample protein content was measured by Coomassie Plus Protein Assay Reagent (Thermo Fisher). For calculation of dose-response curves and $IC_{50}$, the 60 min values were fitted using a four-parameter logistic function using GraphPad Prism 5.0 software.

### 2.3. Confocal microscopy of nuclear translocation of NF-κB

Cells, seeded overnight on cover-slides, were incubated with increasing concentrations of inhibitors, and stimulated with TNFα (2 ng/ml) for 30 min. Cells were fixated with 2% paraformaldehyd, washed, and permeabilized with PBS/Tween®20. Cells were stained using primary antibodies for NF-κB (p65), (rabbit pAb, Santa Cruz, Cat. No.: sc-372, 1:1000) and fluorescence-labeled secondary antibody (Alexa Fluor® 488 goat anti-rabbit IgG (H+L), Life Technologies Cat. No.: 11008, 1:2000) for 30 min. Subsequently, the cell nuclei were stained using DAPI (1:40.000, Life Technologies, Cat. No.: D1306) for 30 min. The cover-slides were embedded with Flouramount G (Invitrogen), dried overnight at 4 °C, and evaluated by confocal microscope (Leica, LSM3).

## 2.4. Inhibition of cytokine release *in vivo* in H5N1 infection model

The highly pathogenic avian H5N1 influenza A virus strain A/Mallard/Bavaria/1/2006 (H5N1, MB1), obtained from the Bavarian Health and Food Safety Authority, Oberschleissheim, Germany was grown in embryonated chicken eggs.

Six to eight-week-old Balb/c mice from the animal breeding facilities at the Friedrich-Loeffler-Institute, Federal Research Institute for Animal Health, Tuebingen, Germany, were anaesthetized by intraperitoneal injection of 150 µl of a ketamine (1%, Sanofi)-rompun (0,2%, Bayer) solution before treatment. Balb/c mice were intranasal infected with avian H5N1 virus A/mallard/Bavaria/1/2006 ($7 \times 10^2$ pfu, i.e. 10-fold MLD50). Mice were i.v. treated with 25 mg/kg VL-01 or solvent (mock) 2 hours prior to virus infection. Serum samples for cytokine analysis were collected before and 12h, 30h, or 72 h after infection. All animal studies were approved by the Institutional Animal Care and Use Committee of Tuebingen.

## 2.5. Inhibition of cytokine release *in vivo* in LPS challenge model

To investigate the effect of VL-01 on LPS induced cytokine response, mice were i.v. treated with 25 mg/kg VL-01 two hours prior to LPS treatment (Lipopolysaccharides from Escherichia coli 055:B5, Sigma, Germany, 20 µg/mice). Serum samples for cytokine analysis were collected before (-4h) and 1.5h and 3h after LPS treatment.

## 2.6. Cytokine analysis

Cytokine analysis was performed using Bio-Plex Protein Arrays from BioRad (Bio-Rad Laboratories, Munich). Bio-Plex-Pro-Mouse Cytokine 6-Plex or 23-Plex were used for cytokine analysis after H5N1 infection or LPS challenge, respectively.

## 3. Results

**3.1. Specific inhibition of chymotrypsin-like activity of the proteasome inhibitor VL-01**

The 26S proteasome comprises three catalytically active subunits β1, β2, and β5 that exhibit caspase-like (Cas), trypsin-like (Try), and chymotrypsin-like (Chy) activities, respectively (Kisselevet al., 2005). To investigate specificity and potency, the new proteasome inhibitor VL-01 (structure shown in Fig. 2A) was tested in comparison to the FDA approved proteasome inhibitor Bortezomib Velcade® in Jurkat T-cell lysates with regard to their inhibitory pattern for different catalytic sites using substrates specific for ß1, ß2, and ß5, respectively. VL-01 inhibited the β5 chymotryptic activity at a 100 fold higher $IC_{50}$ compared to Bortezomib (448 nM vs. 3 nM), with a higher specificity for the ß5 proteolytic site, with low or no inhibition of ß1 caspase activity or ß2 trypsin-like activity, respectively (Fig. 2B,C).

**3.2. Inhibition of NF-κB nuclear translocation *in vitro* after TNFα stimulation by the proteasome inhibitor VL-01**

We investigated whether VL-01 inhibits the translocation of NF-κB dimer (p65/p50) from the cytoplasm to the nucleus. Without stimulation of the NF-κB pathway, p65/p50 is normally sequestered in the cytosol by its inhibitor IκB. Following stimulation by various signal pathways, e.g. TNFα as used in the present study, NF-κB translocates to the nucleus, where it initiates the transcription of responsive genes. In unstimulated Huh7 cells or human monocyte-derived macrophages the p65/p50 dimer is located in the cytosol as demonstrated by merging of nucleus staining (DAPI) and p65 staining (Fig. A,B upper panel; Control). When both cell types were stimulated with TNFα, NF-κB translocated to the nucleus shown by FITC staining (Fig. A,B second panel; 2 ng TNFα). NF-κB nuclear translocation after TNFα stimulation was inhibited by application of 2 μM VL-01 to Huh7 cells (Fig. 3A) and human monocyte-derived macrophages (Fig. 3B) showing p65 staining in the cytosol and only few cells with p65 positive nucleus (Fig. A,B lower panel; VL-01/2 ng TNFα). Application of the proteasome inhibitor VL-01 in the absence of TNFα stimulation had no effect on p65 translocation (Fig. A third panel; 3 μM VL-01).

**3.3. Inhibition of cytokine storm in mice after avian H5N1 Influenza A infection or LPS injection by the proteasome inhibitor VL-01**

The influence of VL-01 on the pro-inflammatory cytokine and chemokine response *in vivo* was tested in a H5N1 influenza virus mouse model. H5N1 infection induces a strong cytokine and chemokine response in mice. Balb/c mice were intranasally infected with avian H5N1 virus A/mallard/Bavaria/1/2006 ($7 \times 10^2$ pfu, i.e. 10-fold $MLD_{50}$). Mice were treated i.v. either with 25 mg/kg VL-01 or solvent (mock) two hours prior to virus infection. Serum samples for cytokine analysis were collected at 0 or 12, 30 or 72 hrs after infection. Different release kinetics of different cytokines and chemokines were found following H5N1 influenza virus infection. While TNFα, MIP-1β, and RANTES peaked very early after H5N1 infection, after 12 hrs, others, *i.e.* KC (neutrophil-activating protein-3) and IL-6, peaked at 72 hrs after infection (Fig. 4). Treatment with VL-01 significantly ($p < 0.05$) inhibited the release of IL-1, IL-6, TNFα, MIP-1 and CXCL1 at the peak time-points in Balb/c mice after infection with the highly pathogenic avian H5N1 influenza A virus (Fig. 4). Importantly, proteasome inhibition significantly decreased the release for *all*, *early and late* cytokines and chemokines in mice infected with avian Influenza A virus H5N1.

Since VL-01 inhibits also H5N1 influenza virus replication in mice (Haasbach et al 2011) one might argue that the observed reduction of cytokines and chemokines could be due to a reduction in viral load rather than a direct influence on the NF-κB signaling itself. Thus, an acute lung injury (ALI) mouse model was used to investigate, whether a similar inhibitory effect of the proteasome inhibitor *in vivo* was found also after LPS challenge. This model provides a rapid and strong systemic induction of pro-inflammatory cytokines and chemokines. Balb/c mice were treated i.v. with 25 mg/kg VL-01, followed by i.p. application of 20 μg LPS. Serum samples for cytokine analysis were collected before (-4hrs) LPS treatment (control) and two time-points after LPS treatment (1.5 and 3 hrs). Again distinct release patterns were found for the different cytokines/chemokines, with TNFα, IL-1β, MIP-1α and MIP-1β peaking already 1.5 hrs after LPS challenge, followed by others, such as IL-6, RANTES, IL-12p40 and KC peaking 3 hrs after LPS stimulus (Fig. 5). Importantly, treatment of mice with proteasome inhibitor significantly ($p < 0.05$) reduced release of the whole panel of pro-inflammatory cytokines and chemokines. **(Fig. 4,5).**

Taken together, these data generated in different models demonstrate the principal potency of VL-01 to interfere with the pro-inflammatory effects, by inhibiting the translocation of NF-κB to the nucleus.

## 4. Discussion

Highly dysregulated exuberant inflammatory responses, with significantly elevated cytokine levels correlate with severity of disease and lethality in COVID-19 patients. Importantly, this cytokine storm seems to be a common hallmark of critical stage COVID-19 patients, but also after H1N1, H5N1, SARS-CoV, or MERS-CoV infection. We could show that the excessive cytokine release induced by highly pathogenic avian H5N1 Influenza A virus was significantly reduced by application of the proteasome inhibitor VL-01 (Fig. 4). A similar inhibition was found for the cytokine cascade induced in mice after LPS injection (Fig. 5). Central to this inhibition is the blocking of the nuclear translocation of NF-κB by VL-01 (Fig. 2). In this line, anti-inflammatory responses have been described in various clinical studies using bortezomib (van der Heijden et al., 2009; Lassoued et al., 2019; Sun et al., 2018; Liu et al., 2016; Jakez-Ocampo et al., 2015).

Reaching beyond the possibilities of currently evaluated drugs for single targets of the cytokine cascade, e.g. monoclonal antibodies against the IL-6 receptor (Moore & June, 2020; Radbel et al., 2020; Aziz et al., 2020; Zhang et al., 2020; Xu et al., 2020) or IL-1 receptor antagonist (Cavalli et al., 2020), inhibition of NF-κB by proteasome inhibitors could provide the unique potential to inhibit the release of multiple cytokines simultaneously, in particular strongly pro-inflammatory cytokines including IL-1, IL-6, TNFα and chemokines including MIP-1 and CXCL1 (KC, neutrophil-activating protein-3).

Several registered proteasome inhibitors (Bortezomib, Carfilzomib or Ixazomib) are available for treatment of multiple myeloma and Mantel cell lymphoma. The drug related adverse events are well known (including thrombocytopenia, neutropenia, peripheral neuropathy (Berenson et al., 2005; Richardson et al., 2006)) which may allow rather lean study protocols for testing these compounds for their therapeutic efficacy in COVID-19 patients.

In contrast to oncological indications where eight (or more) treatment cycles are routinely applied, it seems plausible that just few applications of proteasome inhibitors will be sufficient to downregulate the acute cytokine storm in COVID-19 patients. This may mean better tolerability since the prevalence of side effects seems to correlate with the cumulative dose of bortezomib (Berenson et al., 2005; Richardson et al., 2006). In this line, few applications of the IL-6 receptor-specific antibody tocilizumab were found to be sufficient to trigger a clinical effect in critical COVID-19 patients (Xu et al., 2020).

Importantly, compared to single target approaches, a simultaneous inhibition of multiple cytokines/chemokines using proteasome inhibitors, or alternatively by other inhibitors of the NF-κB pathway, may be highly advantageous compared to single target approaches to compensate for redundant, synergistic, and triggering effects of multiple cytokines (i.e. cytokine cascade) released in critical cases of highly pathogenic hCoV infection (but also H5N1 or H1N1 infection).

This concept is supported by two recently published studies showing pronounced clinical effect in critical COVID-19 patients by Bruton tyrosine kinase (BTK) inhibitors, correlating with significantly decrease in inflammatory parameters (C-reactive protein and IL-6), normalized lymphopenia, and improved oxygenation, with the correlation between BTK and NF-κB-dependent cytokine release discussed (Treon et al., 2020; Roschewski et al., 2020).

Further support for the present concept is provided by recent results from the RECOVERY trial (http://www.ox.ac.uk/news/2020-06-16-low-cost-dexamethasone-reduces-death-one-third-hospitalised-patients-severe). Dexamethasone was found to significantly reduce death in patients with severe respiratory complications of COVID-19 requiring ventilation by up to one third. Dexamethason – a broadly used glucocorticoid anti-inflammatory drug – is assumed to mediate its anti-inflammatory activity at least partially via downregulation of the NF-κB activity (Yamamoto & Gaynor, 2001; Aghai et al, 2006; Meduri et al., 2005).

In contrast to another recently suggested systemic approach for simultaneous inhibition of cytokines by JAK inhibitors (Spinelli et al., 2020), NF-κB inhibition will inhibit predominantly highly pro-inflammatory cytokines and chemokines, such as TNFα, IL-1, IL-6, MCP-1, MIP-1, which are expected to be primarily involved in exuberant systemic inflammatory responses rather than cytokines primarily involved in antiviral responsiveness, such as IFNγ – which is dependent on other pathways, i.e. JAK/STAT.

Although there are still many open questions, the potential to control the cytokine storm-induced severe lung failure and systemic organ failure by using already registered drugs, like proteasome inhibitors, or other inhibitors of the NF-κB pathway may be a real chance to get an additional treatment option, hopefully decreasing the number of cases in need for artificial ventilation and death.

**Acknowledgments**

We thank Dr. Lilija Kircheis for critical reading of the manuscript.

**Declarations of interest:**

none

**Figure Legends**

**Fig. 1: Activation pathway of NF-κB and linkage to virus induced cytokine storm**

Binding of SARS-CoV-2 to its receptor, i.e. the angiotensin-converting enzyme 2 (ACE2) and the help of the cellular serine protease TMPRSS2 trigger endocytosis into the host cell. Within the endosomes, RNA from single-stranded RNA virus is known to activate the Toll-like receptors TLR7/8. This can activate transcription of the interferon-regulator factor (IRF) family and antiviral responses (green dotted lines). However, as a second effect, the activation of the TLR7/8 can trigger - via various intermediates – the activation of IKK (IκB kinases) (grey dotted lines) resulting in phosphorylation of the cytoplasmic inhibitor factor IκBα leading to its degradation by the 26S proteasome, thereby NF-κB (a heterodimer complex of protein subunits p50 and p65) is released from IκBα and can now enter the nucleus and initiate transcription of various genes coding for pro-inflammatory cytokines, chemokines, adhesion molecules, and growth factors. Importantly, this final sequence of NF-κB activation is shared with a variety of cytokine receptor- and Toll-like receptor-mediated signal cascades, including binding of TNFα or IL-1 to their receptors or binding of LPS (e.g. from secondary bacterial infections) to TLR4. In contrast, interferon-response factor (IRF)-related responses are largely independent on NF-κB translocation. Excessive pro-inflammatory cytokine release leads to inflammatory cell activation and infiltration, vascular leakage syndrome, leading to pulmonary oedema and pneumonia. This figure presents a hypothesis from literature-derived information (Hoffmann et al., 2020; Tay et al., 2020; Schett et al., 2020; Moynagh, 2005).

**Fig. 2: Specific inhibition of chymotrypsin-like activity of the proteasome inhibitor VL-01**

VL-01 is highly specific for the chymotrypsin-like activity of the proteasome. Chemical structure of VL-01 (**A**) and analysis of chymotrypsin-like (Chy, ß5 subunit of proteasome), trypsin-like (Try, ß2 subunit) and caspase-like (Cas, ß1 subunit) activities of proteasomes after dose-dependent inhibition with inhibitor VL-01 (**B**) or Bortezomib (**C**) are shown. The effective inhibitory concentration (IC50) is shown in nM.

**Fig. 3: Inhibition of TNFα induced nuclear translocation of NF-κB by the proteasome inhibitor VL-01**

Human Huh7 cells (**A**) or human monocyte-derived macrophages (**B**) seeded on cover-slides overnight, were incubated with increasing concentrations of proteasome inhibitor and stimulated with TNFα (2 ng/ml) for 30 min. Immunofluorescence staining of NF-κB was done using a p65 specific antibody, and cell nuclei were counterstained with DAPI and evaluated by confocal microscope (Leica, LSM3).

**Fig. 4. Inhibition of cytokine release in BALB/c mice infected with H5N1 by treatment with the proteasome inhibitor VL-01**

Balb/c mice (n=4) were intranasal infected with avian H5N1 virus A/mallard/Bavaria/1/2006 ($7 \times 10^2$ pfu, i.e. 10-fold MLD50). Mice were i.v. treated with 25 mg/kg VL-01 or solvent (mock) two hours prior to virus infection. Cytokine levels in blood were determined before (0h) and 12h, 30h or 72 h after infection using the Bio-Plex Pro Mouse Cytokine 6-Plex Panel (Biorad). * $p<0.05$.

**Fig. 5. Inhibition of LPS-induced cytokine release in BALB/c mice by the proteasome inhibitor VL-01**

Balb/c mice (n=7) were injected i.v. with VL-01 (25 mg/kg, in 200µl i.v.) two hours prior LPS stimulation (20 µg/mouse. 200 ml i.p.). Cytokine levels were determined for the time points before (-4h), and 1.5h or 3h after LPS injection, using the Bio-Plex Pro Mouse Cytokine 23-Plex Panel (Biorad). * $p<0.05$

Fig. 1

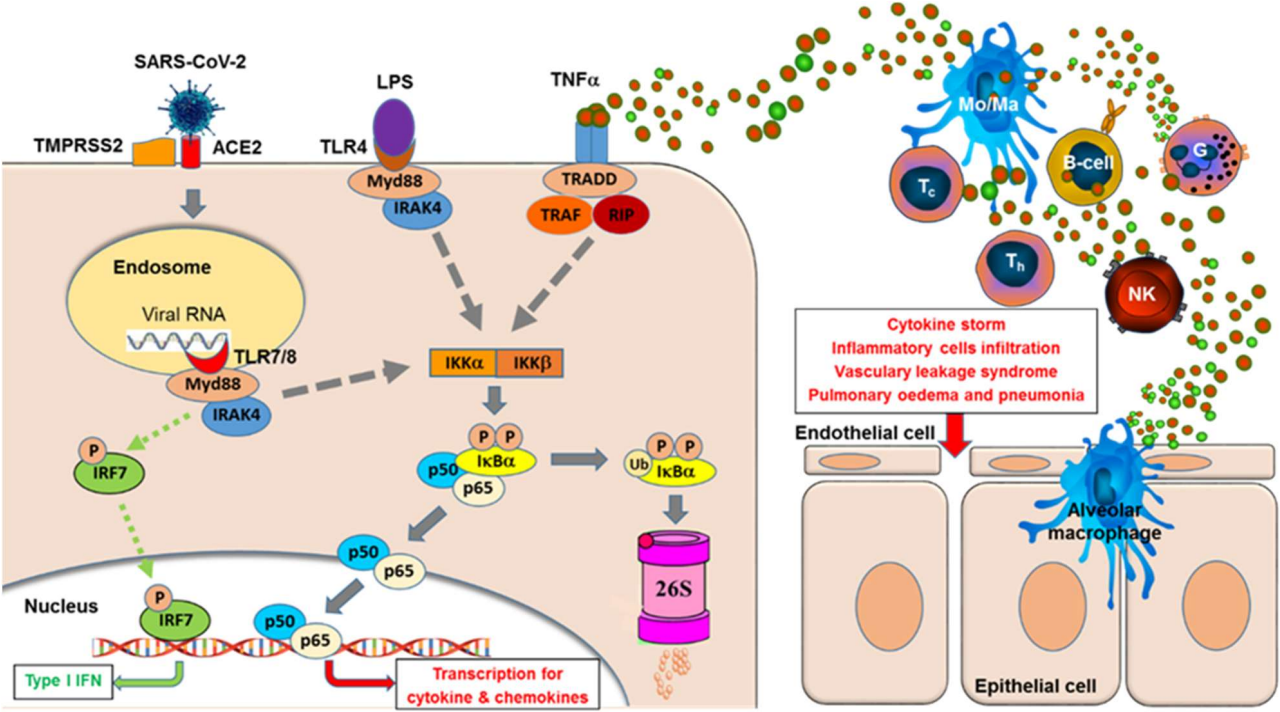

Fig. 2

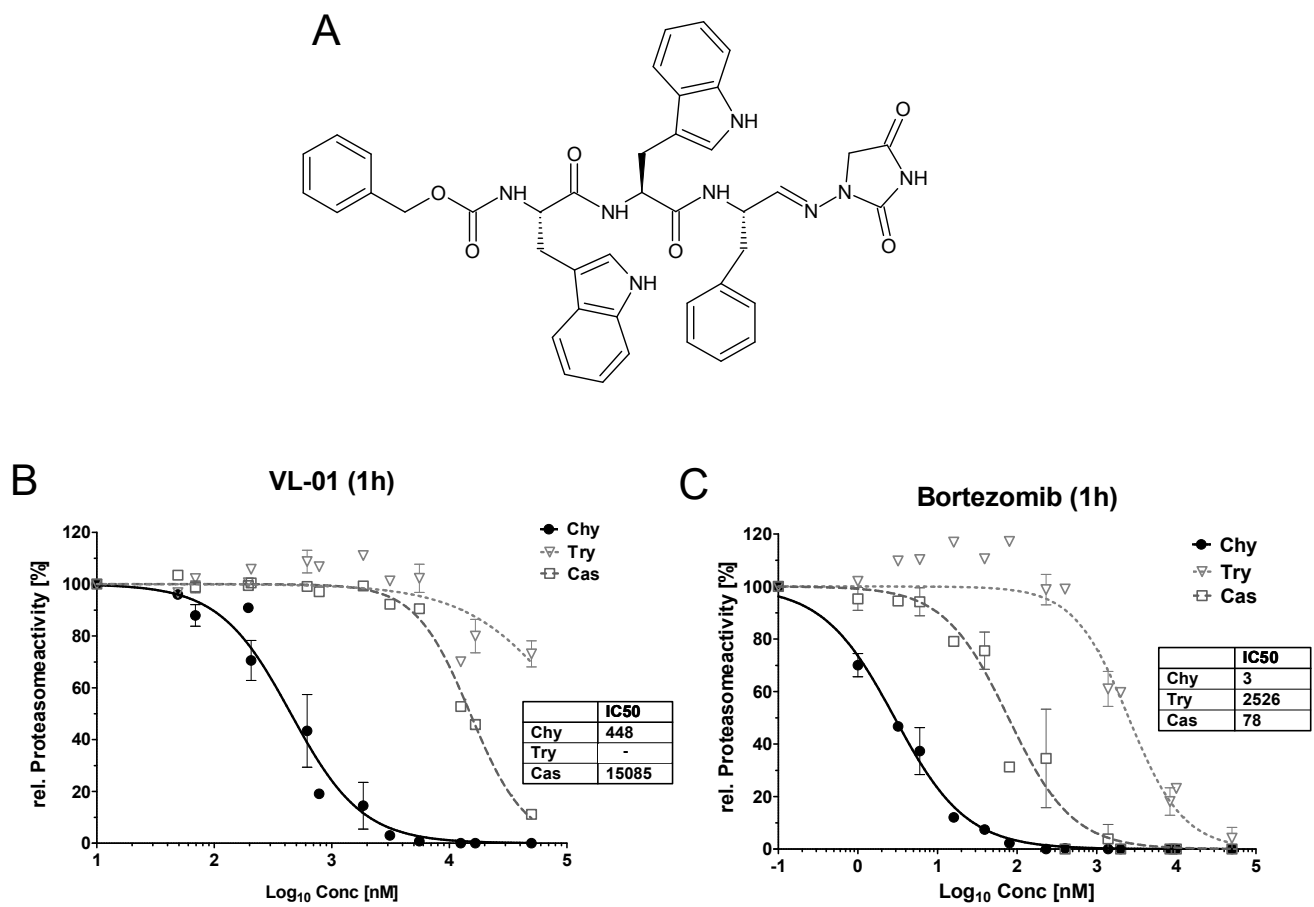

Fig. 3

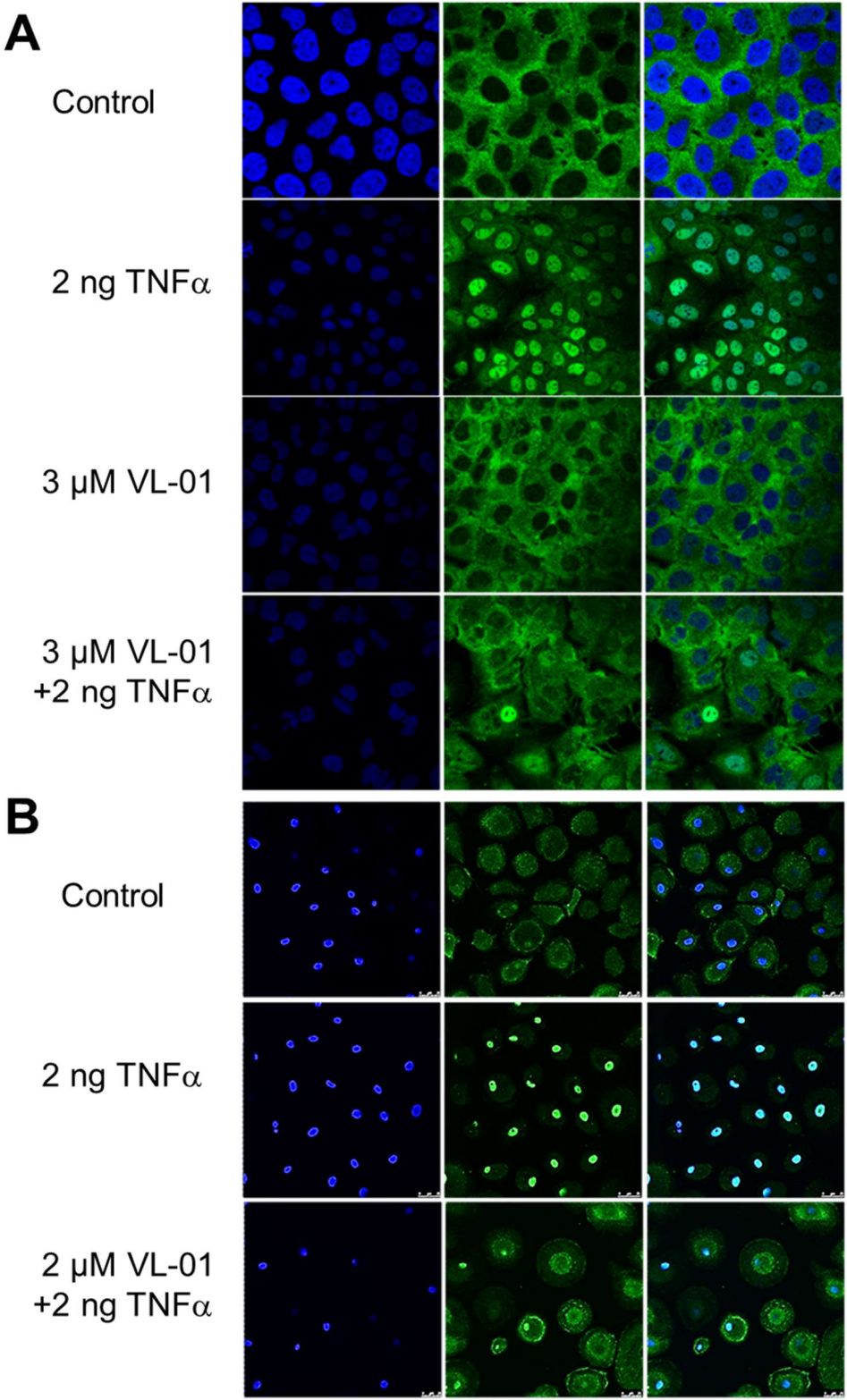

Fig. 4

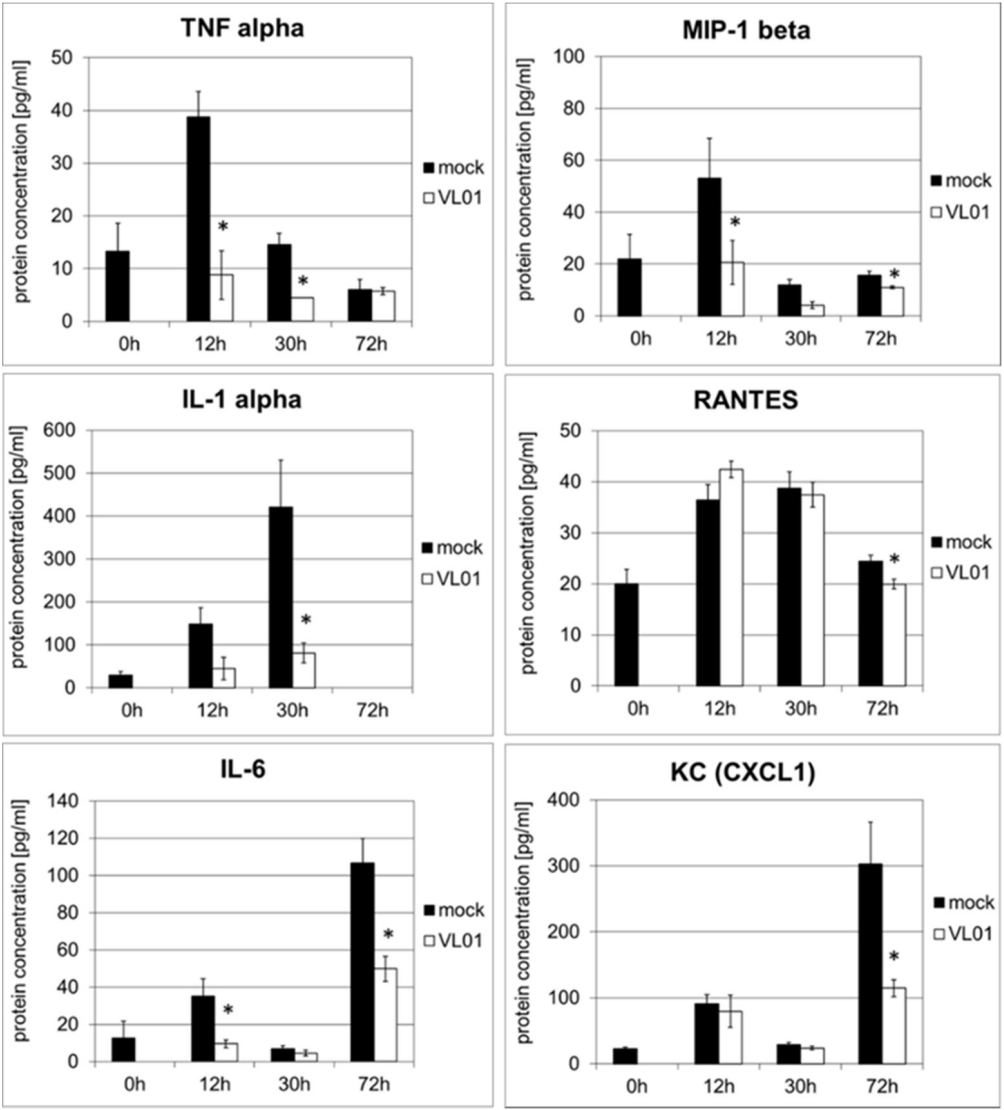

Fig. 5

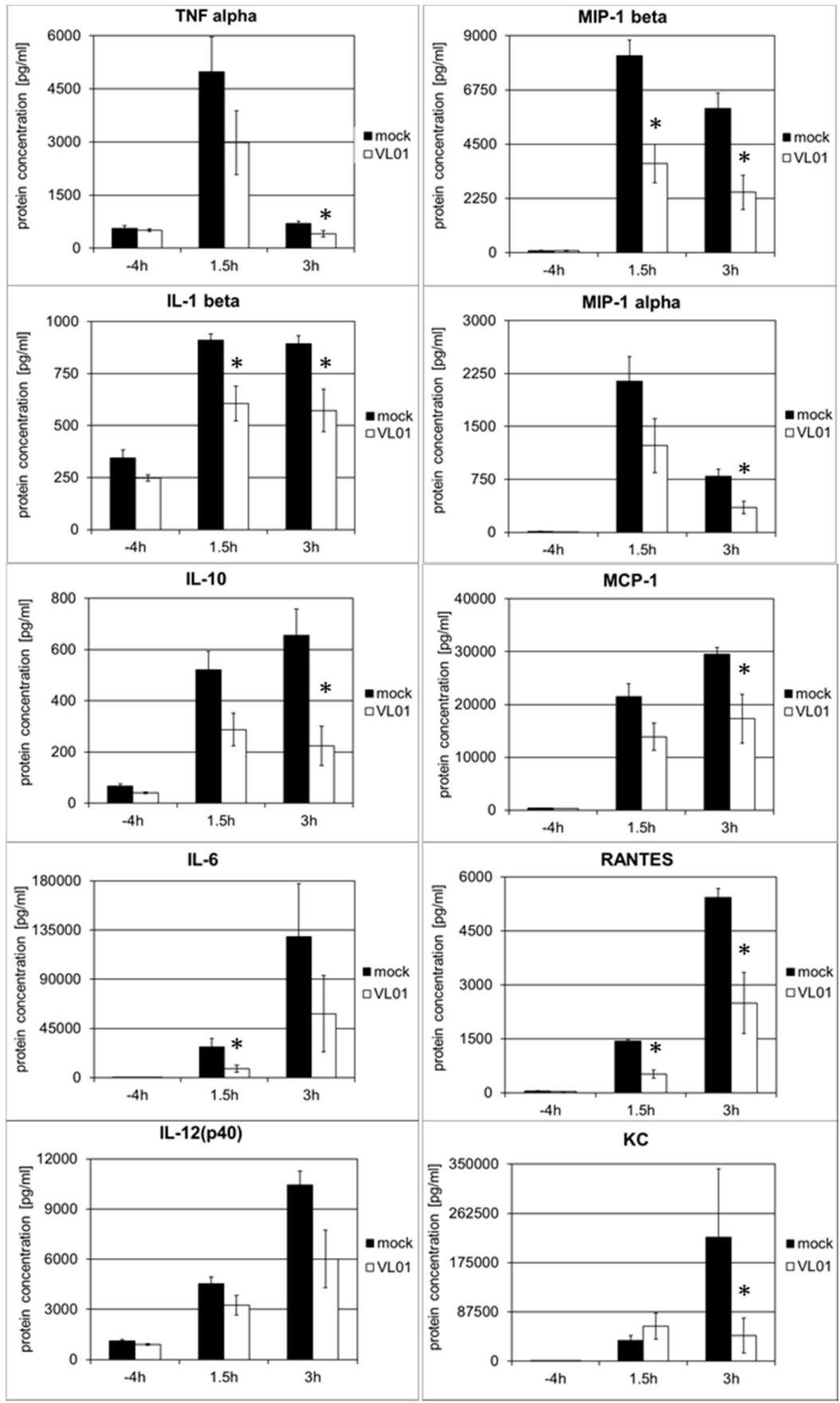